# Energy-resolved EBSD using a monolithic direct electron detector

Nicolò M. Della Ventura [a,*,1], Kalani Moore [b,*,1], McLean P. Echlin [a], Matthew R. Begley [a], Tresa M. Pollock [a], Marc De Graef [c,*], Daniel S. Gianola [a,*]

[a] Materials Department, University of California Santa Barbara, Santa Barbara, CA, USA
[b] Direct Electron L.P., San Diego, CA, USA
[c] Department of Materials Science and Engineering, Carnegie Mellon University, Pittsburgh, PA, USA

ABSTRACT

Accurate quantification of the energy distribution of backscattered electrons (BSEs) contributing to electron backscatter diffraction (EBSD) patterns remains as an active challenge. This study introduces an energy-resolved EBSD methodology based on a monolithic active pixel sensor direct electron detector and an electron-counting algorithm to enable the energy quantification of individual BSEs, providing direct measurements of electron energy spectra within diffraction patterns. Following detector calibration of the detector signal as a function of primary beam energy, measurements using a 12 keV primary beam on Si(100) reveal a broad BSE energy distribution across the diffraction pattern, extending down to 3 keV. Furthermore, an angular dependence in the weighted average BSE energy is observed, closely matching predictions from Monte Carlo simulations. Pixel-resolved energy maps reveal subtle modulations at Kikuchi band edges, offering insights into the backscattering process. By applying energy filtering within spectral windows as narrow as 2 keV centered on the primary beam energy, significant enhancement in pattern clarity and high-frequency detail is observed. Notably, BSEs in the 9–10 keV range dominate Kikuchi pattern formation, while BSEs in the 2–8 keV range, despite having undergone substantial energy loss, still produce Kikuchi patterns. By enabling energy determination at the single-electron level, this approach introduces a versatile tool-set for expanding the quantitative capabilities of EBSD, thereby offering the potential to deepen the understanding of diffraction contrast mechanisms and to advance the precision of crystallographic measurements.

## 1. Introduction

### 1.1. Motivation for energy-resolved EBSD

Electron backscatter diffraction (EBSD) has become an indispensable technique for characterizing crystallographic orientation, phase distribution, and strain in a wide variety of materials [1–4]. Despite its widespread adoption, a long-standing interest within the EBSD community has been the direct quantification of the energy of backscattered electrons (BSEs) contributing to diffraction pattern formation [5–14]. The energy spectrum of BSEs encodes valuable information about the nature of electron–matter interactions; however, this dimension remains largely unexplored in experimental practice. This is particularly important given that multiple scattering processes, both elastic and inelastic, occur for each electron, which introduce energy variation across the diffraction pattern [10,12]. Nevertheless, most theoretical models employed in EBSD simplify this complexity by assuming quasi-elastic scattering with either fixed or average electron energies [6,8]. As a consequence, the role of inelastic losses, the effects of depth-dependent scattering [15], and the potential for energy-resolved contrast mechanisms remain only partially understood.

Access to the energy of each detected electron can improve EBSD by allowing a more detailed analysis of the interaction volume and the mechanisms behind contrast formation. Energy-resolved measurements help distinguish electrons based on their scattering history, which supports more accurate interpretation of diffraction patterns. This information also benefits theoretical modeling and simulations by accounting for energy-dependent scattering. In practice, filtering electrons by energy can improve pattern quality, reduce background noise, isolate near-surface information, and adjust contrast to emphasize specific microstructural features.

Numerous efforts have been made over the past two decades to introduce energy selectivity into EBSD measurements. Initial approaches focused on hardware modifications, such as incorporating energy-filtering electron optics and post-sample lens systems in conjunction with phosphor screens to limit detection to narrow energy bands [7,16]. These methods aimed to enhance contrast by physically restricting the range of detected electron energies [17] but were limited by the performance of the scintillator–CCD/CMOS (charge-coupled devices/complementary metal–oxide–semiconductor) detection chain. With the advent of direct electron detectors (DEDs), and their higher efficiency over their indirect counterparts [18–20], significant advances in energy-filtered EBSD have been achieved, especially through the use of hybrid pixel DEDs such as those based on Medipix technology. Using a discriminator component built into the detector [21,22], electrons above a user-defined energy value are registered, enabling discrete energy thresholding capabilities without modifying the acquisition hardware, improving signal-to-noise ratio (SNR) and thereby enhancing pattern fidelity [23,24]. The most recent advances in this direction involve applying energy filtering during acquisition to entire EBSD maps or diffraction pattern datasets [25], rather than limiting the analysis to individual patterns. Parallel to these hardware innovations, computational techniques also emerged to refine the interpretation of EBSD patterns. For instance, digital image correlation (DIC) has been used to improve angular resolution by comparing experimental data to simulated energy-filtered patterns, offering a virtual means of extracting energy-resolved information [26].

* Corresponding authors.
*E-mail addresses:* n_dellaventura@ucsb.edu (N.M. Della Ventura), kalanimoore@gmail.com (K. Moore), mdg@andrew.cmu.edu (M. De Graef), gianola@ucsb.edu (D.S. Gianola).
[1] These authors share first and corresponding authorship.

As interest in electron energy quantification and energy-filtered EBSD grows, attention has increasingly turned to detector technologies capable of both high energy sensitivity and high spatial resolution. This context has positioned DEDs as a compelling solution, particularly in light of their more recent implementations [19].

### 1.2. Monolithic active pixel sensors (MAPS): Electron counting for energy quantification

Direct electron detectors were originally developed to address the stringent imaging requirements of single-particle cryogenic electron microscopy (Cryo-EM) [27–29], where maximizing detector quantum efficiency (DQE) is paramount. Biological samples used in Cryo-EM are highly sensitive to electron damage, making it essential to capture nearly every scattered electron to achieve high-resolution structure determination. Phosphor-based detectors using a scintillator-coupled to CCD or CMOS sensors suffer from reduced DQE for both low-spatial frequency features, such as broad zone-axis diffraction disks, and high-frequency details such as Kikuchi band edges [22,30,31]. These limitations were overcome with the introduction of monolithic active pixel sensors (MAPS), radiation-hardened silicon devices that allow electrons to directly impact the sensitive layer. MAPS detectors offer excellent DQE across the spatial frequency spectrum, playing a central role in the so-called "resolution revolution" in Cryo-EM, which culminated in the awarding of the 2017 Nobel Prize in Chemistry [32].

Complementing MAPS devices are hybrid pixel detectors, developed primarily for transmission electron microscopy (TEM) diffraction applications [22,33–36]. These detectors employ a thick sensor architecture designed to fully stop incident high-energy electrons within the volume of a single pixel. As a result, the total detected signal correlates directly to the energy of the electron. However, to ensure complete energy deposition within individual pixels, hybrid detectors require relatively large pixel sizes, typically ranging from 75 to 150 μm. This design constraint limits both their spatial resolution and total array size, with most common systems maxing out at 256 × 256 pixels and the largest extending 1,028 × 1,062 pixels [37,38] . In contrast, MAPS detectors feature thin silicon layers through which high-energy electrons (as encountered in TEM) can pass, depositing only a portion of their energy along the way. This partial energy sampling introduces statistical fluctuations in energy deposition known as Landau noise. Landau noise disrupts the direct relationship between deposited signal and the energy of incident electrons, as only a fraction of the electron's energy is stochastically transferred to the detector. This results in a reduction of the detector quantum efficiency at zero spatial frequency, DQE(0), often to values below 0.4 [39]. Despite this, MAPS detectors offer significantly higher spatial resolution due to their fine pixel pitch (6–15 μm) and large array sizes, extending up to 4096 × 4096 pixels (as exemplified by the specific detector employed in this study [19]).

The operational flexibility of MAPS detectors further enhances their utility, as they support two distinct acquisition modalities: integrating and counting. In integrating mode, a MAPS detector accumulates signal in each pixel throughout the duration of a frame. The intensity of each electron interaction is registered as an arbitrary digital unit (ADU) value that represents the total charge deposited in that pixel. As a result, in TEM, this mode is affected by Landau noise, reducing the DQE. To mitigate the loss in DQE in TEM, acquisition in electron-counting mode was introduced. In this modality, a dedicated counting algorithm processes each frame to identify localized clusters of non-zero ADU pixels, sum their ADU values, determine their centroids, and assign a discrete value of 1 to the event's centroid position, regardless of the deposited energy. Each charge is hence registered as a binary event; that is, all electrons are assigned equal weight in the image reconstruction process. The signal is thus not affected by Landau noise and consequently, in TEM, it improves the DQE(0) above 0.95. This acquisition mode is analogous to the operational principle of hybrid pixel detectors when used in Medipix mode, where the sensor functions as a digital counter [23].

The transition from TEM to scanning electron microscopy (SEM), where lower-energy primary beam electrons are used, fundamentally alters the performance landscape. In particular, at low accelerating voltages (below 25 keV for the DE-SEMCam detector used in this study, see Fig. A.1 in Appendix A), incident electrons are almost entirely absorbed within the thin active layer of the MAPS detector, delivering a nearly uniform and complete energy dose to the detection volume. Consequently, image contrast in SEM can be modulated through the energy-dependent weighting of individual electron events, as Landau noise ceases to be a limiting factor in signal detection. This stands in contrast to TEM, where precise quantification of the energy deposited by each electron, achieved via a counting algorithm (e.g., [39]), produces a characteristic Landau-shaped energy spectrum. In SEM, however, the corresponding spectrum ideally features a narrow, single peak, enabling accurate and reproducible energy discrimination for each detected event. In essence, the MAPS detector, when operated in integrating mode within the SEM, achieves performance comparable to that of a hybrid pixel detector, while offering superior spatial resolution. The high spatial fidelity of the MAPS detector makes it particularly well-suited for capturing high-quality EBSD patterns [19,40–42], including those from beam-sensitive materials [43]. In the primary beam energy range of the SEM, operating in electron-counting mode does not provide additional improvement in DQE(0).

To avoid ambiguity, it is essential to clarify the terminology surrounding "counting mode". Typically, *counting mode* denotes an acquisition modality in which a *counting algorithm* is employed during data collection. However, the counting algorithm is a processing method that can also be retro-actively applied to data sets originally acquired in integrating mode to construct the spectrum of electron energy recorded during acquisition. Indeed, following the clustering function, the counting algorithm [39] allows one to measure the energy of each detected electron by converting its ADU signal into an energy value (in keV) using a predetermined ADU/keV calibration factor. It is important to distinguish this capability, i.e., *energy measurement*, from *energy filtering*. Energy measurement refers to assigning an energy value to each detected event, while energy filtering involves selecting only those events whose energies fall within a specified range during acquisition or processing. In energy filtering, the process is further refined to include only those electrons (or better, charge clusters) whose energies lie within a specified range. A binary energy map is hence generated: pixels are assigned a value of 1 if an event within the selected energy window is detected, and 0 otherwise. This energy-based windowing relies on prior energy measurements, allowing for energy filtering at the single-electron level.

In this study, we introduce a novel implementation of energy-resolved EBSD using a MAPS-based DED operating within a conventional SEM environment. We begin by characterizing the spectrum of electron energies across the detector in the absence of a sample to establish the ADU-to-keV calibration. Subsequently, we examine spatial variations in electron energy across Kikuchi patterns acquired with a sample in place. Finally, we demonstrate how targeted energy filtering and different acquisition modalities can enhance the diffraction contrast.

## 2. Methods

### 2.1. Sample preparation

A silicon wafer was purchased from MSE Supplies (Tucson, AZ, USA) and cleaved into a triangular sample of approximately 10 mm edge length to facilitate easy insertion into the SEM. Silver paste was applied to the surface around the scanned area to reduce charging. The sample was mounted on a 70° pre-tilted stub for imaging and EBSD.

### 2.2. Hardware and acquisition methods

Direct electron detection was carried out using the DE-SEMCam manufactured by Direct Electron LP (San Diego, CA), equipped with a custom monolithic active pixel sensor (MAPS, full-frame resolution 4,096 × 4,096, 2× hardware binning 2,048 × 2,048, effective pixel size of 13 μm, maximum readout speed 281 fps) [19]. The DE-SEMCam was installed on a Thermo Fisher Scientific Apreo-S SEM. The microscope was operated at accelerating voltages from 7 to 15 keV. Dark reference backgrounds were collected with the detector in position in the chamber with the electron beam blanked and all photon sources inactive.

Two acquisition modalities were employed in this study: integrating mode and counting mode, described in Section 1.2. In all cases, a sparse signal was acquired, as detailed in Section 2.3. Integrating mode was used for both the determination of the ADU/keV calibration factor (see Section 2.4) and the acquisition of energy-resolved EBSD datasets. A centroid-based counting algorithm [39] was retro-actively applied to these data sets to reconstruct the spectrum of electron energy recorded during acquisition. Acquisition performed directly in counting mode was used to enable real-time energy filtering by registering individual electrons and retaining only those that fall within a specified energy window. This approach allows for precise control over the total collected signal (expressed as electrons per pixel) during data acquisition, thereby ensuring fair comparisons across different energy-filtered conditions.

### 2.3. Parameters for accurate electron counting and energy filtering

A critical requirement for the electron counting algorithm to yield accurate results is to ensure a sufficiently sparse signal, thereby allowing individual electron events to remain isolated. In other words, each pixel containing a detected electron must be surrounded by pixels without a detected signal to ensure that no two adjacent electrons are erroneously registered as a single event. This isolation prevents misidentification due to signal overlap in time and space. This phenomenon, known as coincidence loss, is quantified as the fraction of missed electrons (1 - number of electrons counted/number of incident electrons). An increasing coincidence loss corresponds to a drop in DQE(0). The practical threshold for minimizing this error is approximately one electron per 20 pixels, ensuring a mean sparsity of no more than 0.05 electrons per pixel per frame. Therefore, accurate electron counting with MAPS detectors requires low beam currents and high frame rates to preserve sufficient pixel isolation and avoid signal overlap.

The DED records signal intensities in ADUs, proportional to the energy deposited by each electron. A centroid-based counting algorithm identifies clusters of contiguous non-zero ADU valued pixels corresponding to individual electron events, sums their values to compute the total ADU deposited per event (see Fig. 1a), and assigns this total to the cluster's centroid, denoting the estimated point of electron impact. When standard electron counting acquisition mode is employed, this total energy information is typically discarded, with each event reduced to a binary value of 1. For energy-filtered counting acquisitions, however, the algorithm introduces an additional intermediate step: a predefined energy window (defined by the user in the DE-SEMCam software by specifying a minimum and maximum ADU window) is applied during acquisition such that only electron events whose total ADU values fall within this range are retained and assigned a value of 1; all others are excluded from the final image by assigning a value of 0. To enable this, the approach requires specifying a conversion factor between ADU and keV, determined through the procedure outlined in the following section.

### 2.4. Energy calibration: ADU to keV

As a preliminary step for energy quantification, a calibration has to be performed to determine the ADU/keV conversion factor, i.e., the number of ADUs corresponding to 1 keV of energy deposited by incident electrons. Taking advantage of our detector's four positional degrees of freedom [19], the calibration procedure was carried out with the detector configured in direct-beam mode, with the detector positioned flat such that the electron beam directly impinged on the sensor without any sample (Fig. 1a). To achieve high temporal resolution, the readout area was constrained to 512 × 64 pixels, enabling a frame rate of approximately 14,000 fps. The electron beam dwell time was minimized to 100 ns, the column optics were over-focused to a working distance (WD) of 0 mm, the horizontal field width set to 2.02 mm, and the camera length $L$ was maximized to 44 mm to ensure a highly distributed scan (minimum electron density per solid angle). These parameters (Table 1) were chosen so that, as the beam rastered across the reduced read-out region of the detector, the primary beam remained sufficiently sparse within each frame. For each primary beam energy examined (ranging from 7 to 15 keV in 1 keV increments), a stack of 10,000 frames was acquired in integrating mode using a current of 0.78 pA. During post-processing, a custom centroid-based counting algorithm [39] was applied to each frame within the dataset, producing histograms of the ADU values corresponding to each detected electron, thereby enabling the construction of the detector's energy response profile. For each primary beam energy, the weighted average ADU value of the histogram in Fig. 2 was used to define the ADU/keV conversion factor. This conversion factor is then used to convert from an electron ADU histogram to an energy spectrum (see inset in Fig. 2). The detector energy resolution is quantified by the full-width at half maximum (FWHM). For energy spectra measured from Kikuchi patterns (see Section 3.2), the weighted-average BSE energy ($\overline{E}$) was determined by calculating the intensity-weighted mean of the spectrum.

### 2.5. Simulations

Simulations were carried out using the EMMCOpenCL, EMEBSD-master, and EMEBSD programs from the open source EMsoftOO software package [45] for the same detector geometry as the experiment. The EMEBSD program was configured in a mode that only simulates the continuous background without including dynamical electron scattering. A Monte Carlo (MC) continuous slowing down approximation (CSDA) was used to generate the spatial and energy distributions of 12 keV electrons with an energy bin size of 0.25 keV in the range 3–12 keV. A sample tilt of 70° was used with a total of $16 \times 10^9$ incident electrons and a maximum penetration depth of 100 nm. Individual EBSPs were then computed for the full-size detector and for each energy bin to obtain the spatial and energy distributions of back-scattered electrons across the detector surface.

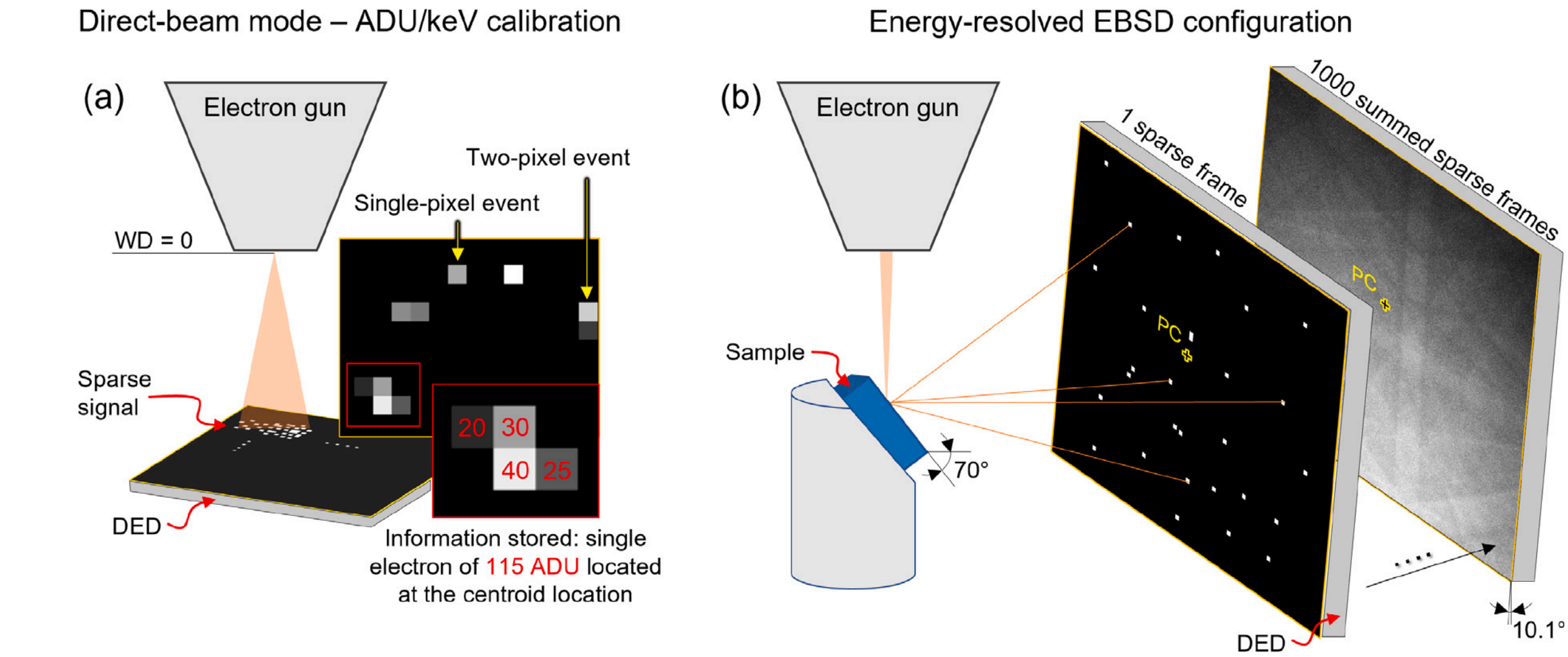

**Fig. 1.** Schematic illustration of the detector configurations used for (a) energy calibration (ADU-to-keV) in direct-beam mode and (b) energy-resolved EBSD pattern acquisition. (a) For calibration, the detector was positioned flat beneath the electron beam (orthogonal to electron beam, 90° detector tilt) to directly measure energy deposition from primary electrons in the absence of a sample. (b) For EBSD measurements, the sample was tilted to 70° and the detector is placed in a conventional geometry (10.1° detector tilt) to capture backscattered electrons exiting the sample surface. In both configurations, data were collected under sparse signal conditions to ensure isolated electron events within each frame, enabling precise single-electron counting. Panel (a) also depicts examples of a single-pixel event (ideal case) and multi-pixel events (non-ideal). In the latter case, the centroid-based counting algorithm aggregates the ADU signal across the pixel cluster and assigns it to the computed centroid position.

**Table 1**
SEM and detector parameters employed in direct-beam mode (Fig. 1a).

| SEM (TFS Apreo-S) Parameters | | | | | |
|---|---|---|---|---|---|
| Operation Mode | HFW (mm) | WD (mm) | Voltage (keV) | Beam Current (pA) | Scan Speed (ns) |
| Standard | 2.02 | 0 | 7–15 | 0.78 | 100 |
| **Detector (DE-SEMCam) Parameters** | | | | | |
| Size (px) | Readout area (px) | Acquisition Mode | Fps | Recorded Frames | Camera Length (mm) |
| 2048 × 2048 | 512 × 64 | Integrating | 14,286 | 10,000 | 44 |

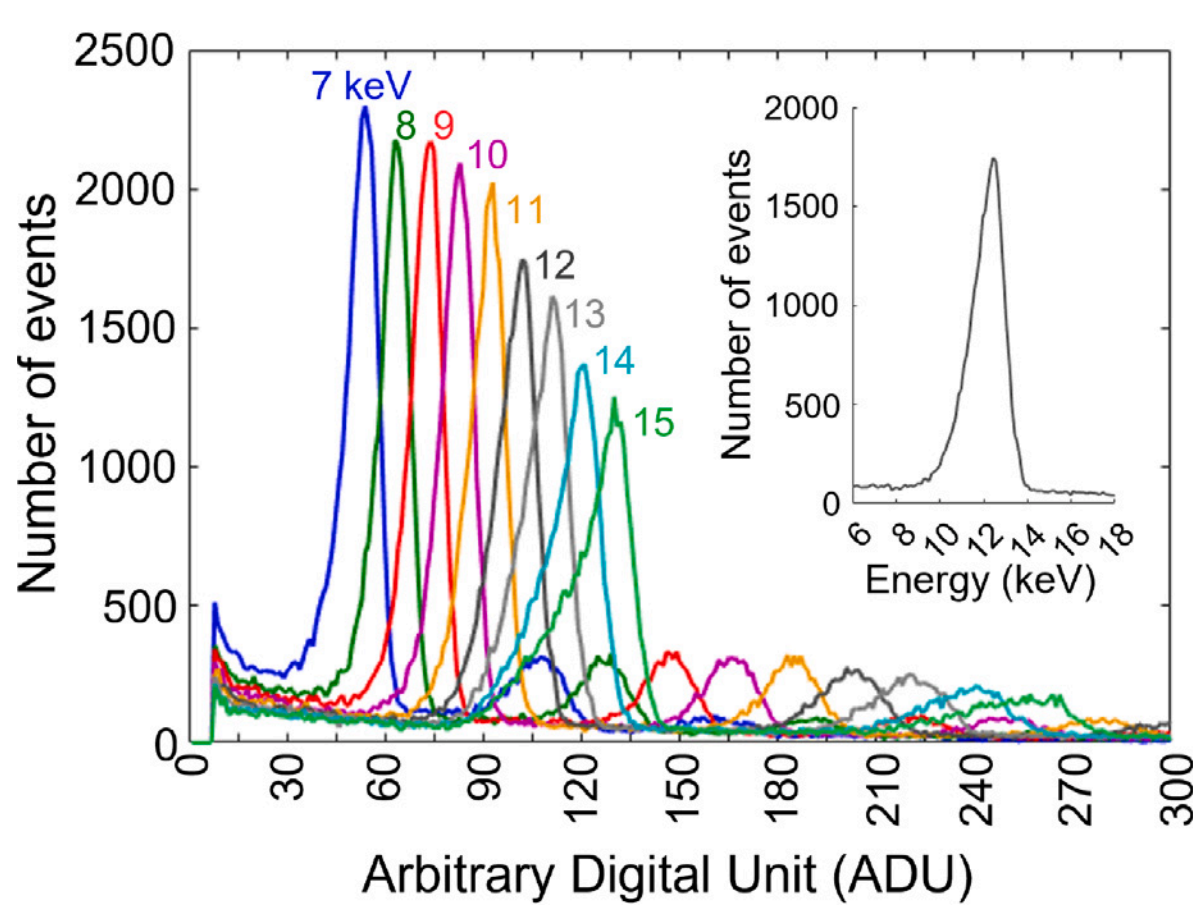

**Fig. 2.** Histograms of electron signal (ADU) measured at accelerating voltages from 7–15 keV. While the $\Delta$Energy of the SEM gun is < 0.001 keV [44], the energy resolution of the DED is ~1 keV, resulting in relatively broad, but distinguishable peaks at each accelerating voltage. Inset: histogram of 12 keV electrons converted to an energy spectrum.

**Table 2**
ADU/keV calibration values.

| | 8 keV | 10 keV | 12 keV | 14 keV |
|---|---|---|---|---|
| ADU/keV | 7.84 | 8.11 | 8.12 | 8.35 |

## 3. Results

### 3.1. Energy calibration: ADU to keV

Fig. 2 presents the histograms of detected ADU values corresponding to primary beam energies ranging from 7 to 15 keV. Each distribution exhibits a well-defined primary peak, corresponding to the most frequent energy deposition per incident electron. The onset of intensity in each histogram (of 8 ADU) reflects the applied intensity threshold used to distinguish the signal from background noise (and other hardware-related artifacts), effectively isolating the localized electron-induced intensity profile. The mean spread of the electron signal on the detector was 1.4 pixels (see Figure S.2 in the Supplementary Materials), reflecting a well-confined point spread function (PSF), and hence modulation transfer function (MTF)[2]. With increasing primary beam energy, the position of the primary peak shifts toward higher ADU values, reflecting the greater energy deposited. For a given voltage, secondary peaks corresponding to quantized two-, three-, and four-electron coincidence events are registered, each of progressively lower intensity. These peaks arise from electrons not spatially separated on the detector and therefore incorrectly registered as a single event, resulting in apparent energy values corresponding to two, three, or four times that of a single electron. The distinct separation of these peaks and the dominance of the primary peak confirm that the measurements were conducted under sufficiently sparse conditions to enable accurate ADU-per-electron calibration.

[2] The PSF characterizes the spatial extent over which the signal from a single point is distributed across adjacent pixels. A broader PSF leads to a degradation in the MTF, which quantifies the camera's ability to preserve edge sharpness and fine detail. High MTF performance corresponds to a high DQE at elevated spatial frequencies, enabling the resolution of fine features (such as band edge sharpness) with reduced signal requirements.

For an accelerating voltage of 12 keV, a conversion factor of 8.12 was determined and subsequently used to re-plot the distribution of electron events as a function of electron energy, as shown in the inset of Fig. 2. The ADU/keV calibration was performed separately for each accelerating voltage as detailed in Section 2.4. The resulting values, some of which are summarized in Table 2, appear very consistent and of approximately 8 ADU/keV, yet show a slight increase with increasing accelerating voltage. This small deviation indicates a minor non-linearity between the accelerating voltage and the ADU peak position. Clear evidence of this can also be observed in the mismatch between the position of the second peak of the 7 keV distribution and that of the first peak of the 14 keV distribution. This non-linearity is likely attributed to energy loss at the sensor surface before electrons reach the sensitive layer, resulting in a small but relatively constant ADU loss per electron, which has a greater effect at lower electron energies. This non-linearity is consistent with observations made during calibration on a Hitachi SEM using the exact same DE-SEMCam [46].

The detector's energy resolution, as quantified by the FWHM of the peaks, is approximately 1 keV. This spread reflects the intrinsic energy resolution of the DED, not of the SEM primary beam energy. Indeed, the SEM beam itself exhibits far superior energy fidelity, with an energy spread at the gun of less than 0.001 keV [44]. Were a 1 keV energy spread intrinsic to the SEM beam, chromatic aberrations would render image formation practically unfeasible. The limited resolution observed in the detector is likely attributable to variability in the generation and collection of electron–hole pairs within the epitaxial layer, as well as their transfer to the floating diffusion layer[3]. Notably, the peak electron count in Fig. 2 diminishes with increasing accelerating voltage, indicating a broadening of the detected signal. This modest reduction in energy resolution likely stems from an increased spatial distribution of electron–hole pairs generated in the epitaxial layer, leading to reduced charge collection efficiency and greater variation in the recorded signal. Again, this observation, along with the 1 keV energy resolution, is consistent with measurements obtained during calibration on a Hitachi SEM equipped with the same DE-SEMCam [46].

### 3.2. Energy-resolved EBSD

After calibrating the keV/ADU ratio specific to the microscope and detector employed, diffraction patterns of Si(100) were acquired to examine the energy distribution of electrons reaching the detector in standard EBSD geometry (Fig. 1b). For an accelerating voltage of 12 keV, a total of 1000 sparse frames (each 2,048 × 2,048 in size) were acquired in integrating mode from an individual point on the sample using the beam scanner in spot mode. Subsequently, these patterns were summed to yield a composite image of the EBSD pattern. An acceleration voltage of 12 keV was selected as it is in the DE-SEMCam's optimal collection efficiency range, which spans from 8 keV to 16 keV [19]. To ensure sparsity of the signal within individual frames, the acquisition was performed at the maximum frame rate of 281 fps using a low beam current of 50 pA, resulting in an average of fewer than 0.010 electrons captured per pixel per frame (see Table 3). Further details of the acquisition parameters are reported in Table 3.

An example frame acquired at 12 keV from Si(100) is shown in Fig. 3a, with a magnified region in Fig. 3b demonstrating the sparse distribution of individual electron events. Selected ADU values are annotated to demonstrate the variability among events. Fig. 3c presents the accumulated ADU intensity map obtained by summing 1000 such sparse frames at full 2,048 × 2,048 resolution. This map reveals the Si(100) diffraction pattern and, when converted via the calibrated ADU/keV factor, serves as a direct representation of the total electron energy deposited at each pixel.

Similar to the calibration procedure, a centroid-based electron counting algorithm was applied to the complete stack of 1000 sparse frames to construct histograms of electron events as a function of ADU values within designated regions of interest (ROIs) across the detector plane. The resulting histograms were subsequently converted from ADU to energy units using the pre-determined ADU/keV calibration parameter. Multiple computational strategies may be employed to define these ROIs, thereby enabling a nuanced interrogation of the spatial distribution of electron energies. In the present study, two complementary methodologies were adopted: (i) a linear analysis of energy distributions along horizontal pixel rows, conducted using five ROIs, each 2,048 × 64 pixels in size and evenly spaced vertically across the detector at the locations marked by the colored bands in Fig. 4a; (ii) a spatial assessment of local energy distributions across the entire detector, performed by uniformly applying an 8 × 8 pixel kernel.

#### 3.2.1. Linear BSE energy profiles across the detector

Fig. 4a presents the same map illustrated in Fig. 3c, but without color coding. Fig. 4b displays the BSE energy spectra extracted from the five ROIs indicated in Fig. 4a, while their corresponding weighted average (w.a.) BSE energy, $\overline{E}$, is reported in Fig. 4c. Near the lower edge of the detector, $\overline{E}$ reaches a mean value of 9.25 keV. Upward along the detector, $\overline{E}$ gradually shifts toward lower values. Near the top of the detector, $\overline{E}$ falls to around 7.71 keV. Importantly, a wide range of BSE energies contribute to the mean value across the entire detector. For example, the energy spectrum exhibits a significant contribution from BSEs with energies markedly lower than that of the incident beam (Fig. 4b). Notably, the peak electron count is observed between 9 and 10 keV, with the value progressively increasing toward the bottom of the detector. These energies, representing roughly 75%–85% of the primary beam energy, indicate that the electrons most active in pattern formation have undergone significant energy loss during their interaction with the material and throughout the diffraction process.

Both the energy and spatial distributions of BSEs were simulated using the MC procedures, incorporating the exact experimental conditions (Table 3) into the computational framework [10], as described in Section 2.5. The simulated cumulative distribution of all BSE energies reaching the scintillator is depicted as a grayscale intensity map in Fig. 5a. The corresponding energy distributions within the selected ROIs are illustrated in Fig. 5b, from which $\overline{E}$ has been extracted and plotted in Fig. 5c. Since the simulations were constrained to the 3–12 keV BSE energy range, the $\overline{E}$ plotted in Fig. 5c were correspondingly extracted from this same interval, and hence referred to as to $\overline{E}_{3-12}$. Close agreement in $\overline{E}_{3-12}$ was observed between the experimental measurements and the simulation results. Nevertheless, differences in the detailed spectral shapes are evident and arise from several factors. In particular, a major contribution originates from the use of the CSDA in the Monte Carlo simulations, which neglects discrete inelastic scattering events and therefore underestimates the near-elastic (zero-loss) peak while overemphasizing the high-loss tails. Similar discrepancies intrinsic to the CSDA framework have been reported previously by Deal et al. [7]. Additional deviations are expected because the simulated patterns represent only the smooth BSE background, whereas the experimental data also include diffraction contrast that modulates the spectral intensity through energy-dependent Kikuchi features. Achieving a one-to-one correspondence between experiment and simulation would require coupling electron transport with a full dynamical diffraction treatment, which lies beyond the present scope. The comparison in Fig. 5b is therefore intended to demonstrate first-order consistency between experiment and simulation and to highlight how the energy-resolved EBSD approach now enables such quantitative analyses and comparisons to be pursued in the future.

[3] This is also consistent with the fact that the maximum ADU value measured in a signal-free frame after dark reference correction is 8 ADU (here not shown), which corresponds to 1 keV attributable solely to noise. Moreover, Fig. 2 reveals a rise in electron counts below 15 ADU, again attributable solely to noise contributions.

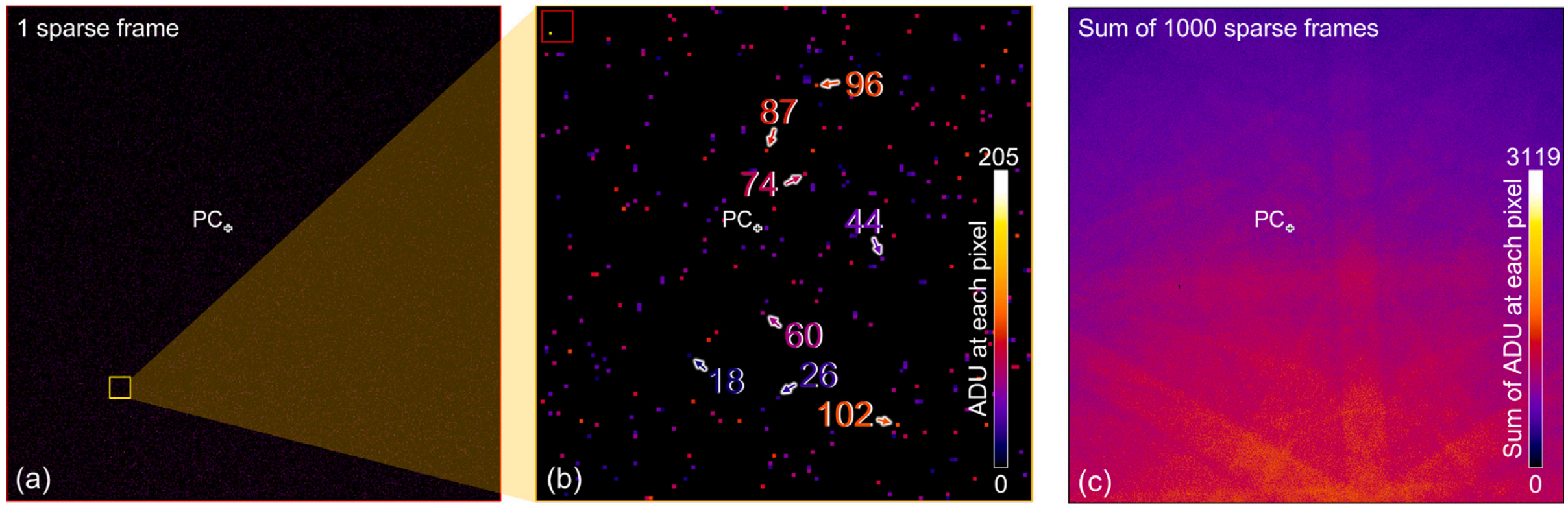

**Fig. 3.** (a) Single frame acquired in Si(100) at an accelerating voltage of 12 keV and a beam current of 50 pA. (b) Magnified region of the single frame showing a sparse distribution of single electrons detected by our sensor. Some of the ADU values are indicated in (b). (c) ADU intensity map at 2,048 × 2,048 pixel resolution, generated by summing pixel-wise ADU values over 1000 sparse frames and revealing a Si(100) diffraction pattern. Using the ADU/keV conversion factor, (c) effectively provides a map of the total electron energy deposited at each pixel, integrated over all frames.

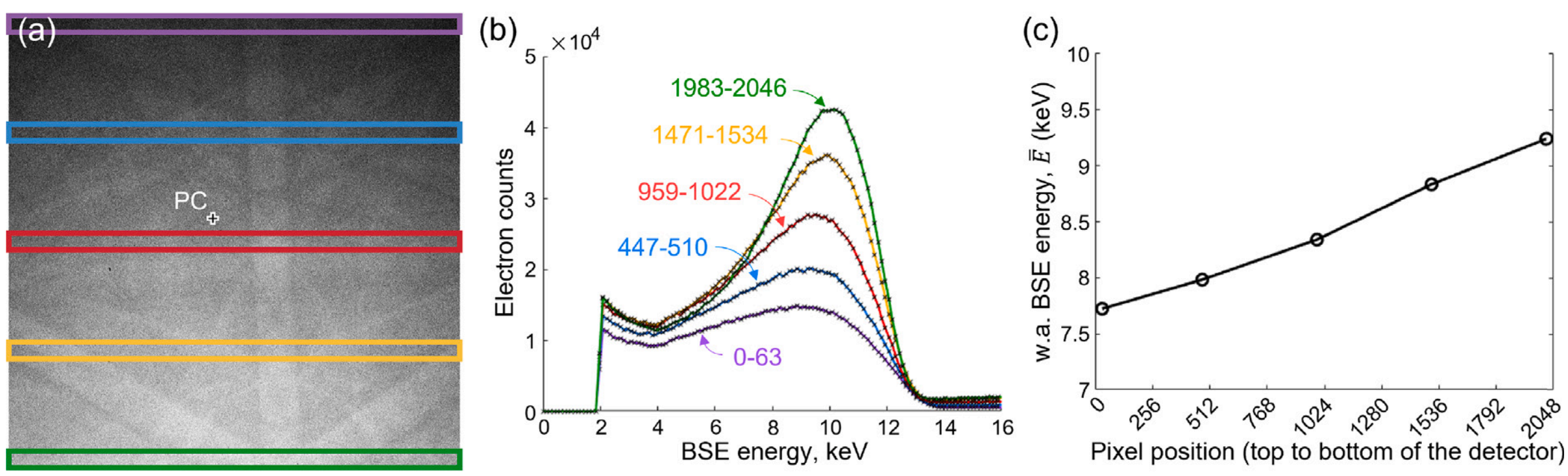

**Fig. 4.** (a) 12 keV Si(100) intensity map obtained by summing 1000 sparsely acquired frames (replicated from Fig. 3c, but without color coding). (b) Electron energy spectra extracted from five vertically spaced ROIs (2,048 px × 64 px) across the detector, illustrating spatial variations in BSE energy distribution. Despite the 12 keV primary beam energy, the measured electron energy distribution exhibits non-zero counts from 2 to 13 keV. The value of 13 keV is consistent with the detector's energy resolution of approximately 1 keV, as discussed in Section 3.1. (c) Weighted average (w.a.) BSE energy, $\overline{E}$, for each ROI, showing a gradual increase from the top to the bottom of the detector.

**Table 3**
Detection parameters for the Si EBSD pattern acquired in spot mode.

| Beam energy (keV) | Beam current (pA) | Fps | Recorded sparse frames | Pattern size | Mean e⁻/px/frame |
|---|---|---|---|---|---|
| 12 | 50 | 281 | 1’000 | 2048 × 2048 | 0.0088 |

| Camera tilt | Working distance, WD (mm) | Camera length, L (mm) | Solid angle | PC (x*, y*, z*) |
|---|---|---|---|---|
| 10.1° | 9.5 | 35.930 | ~41° | 0.4496, 0.5512, 1.3495 |

#### *3.2.2. Spatial distribution of BSE energy on the detector*

We next interrogate the 12 keV dataset by uniformly applying an 8 × 8 pixel kernel across the detector, yielding an 8× spatially binned data cube with dimensions 256 × 256 pixels × 4096 (energy spectra). Fig. 6a shows the resulting 256 × 256 heatmap, representing the $\overline{E}$ values derived from the BSE energy spectrum associated with each kernel.

In addition to the vertically increasing trend in $\overline{E}$ from top to bottom (consistent with the observations discussed previously) a clear modulation in energy emerges at the edges of Kikuchi bands. Concentrating on the prominent vertical $(\overline{1}10)$ Kikuchi band in Fig. 6a, we extract spatial line profiles of the $\overline{E}$ along the trajectories indicated by blue and dark-orange arrows. These profiles, obtained by averaging over detector rows 39–59 (top region) and 215–235 (bottom region), reveal systematic variations in the local energy distribution across the Kikuchi band edges (Figs. 6b and c). Although the mean $\overline{E}$ for these two profiles are approximately 7.92 keV and 9.06 keV, respectively, their spatial behavior is more complex. Specifically, moving laterally from the center of the Kikuchi band (defined as pixel zero) toward the band edge, the $\overline{E}$ remains nearly constant with only minor fluctuations. However, beyond a certain pixel, a distinct decrease in $\overline{E}$ is observed, gradual in the top region (Fig. 6b) and more abrupt in the bottom region (Fig. 6c), highlighting an asymmetry in the spatial decay of $\overline{E}$.

One might initially attribute the observed intensity variation to differences in Kikuchi band width across the detector. However, for a fixed acceleration voltage, bands in the top region are expected to appear narrower than those in the bottom region. This follows from their spatial position relative to the pattern center, as detailed in Section S.1 the Supplementary Materials, where band width is calculated with respect to the pattern center location. Yet, this expectation contradicts the trends observed in Figs. 6b and c. Most importantly, the electron energy inferred from spatial separation between pixels does not necessarily correspond to the $\overline{E}$ derived from the local electron energy spectra at those same pixels. For instance, in the top region (Fig. 6b), the 28-pixel distance between the two points of minimum $\overline{E}$ across the band would imply an electron energy of 6.44 keV to yield such Kikuchi band width, yet the measured energy at those pixels is 7.8 keV. In other words, Fig. 6a should not be interpreted as a conventional EBSP but as a map that reflects energy modulations within the EBSP resulting from a subtle interaction of electrons with the same lattice planes that produce Kikuchi bands.

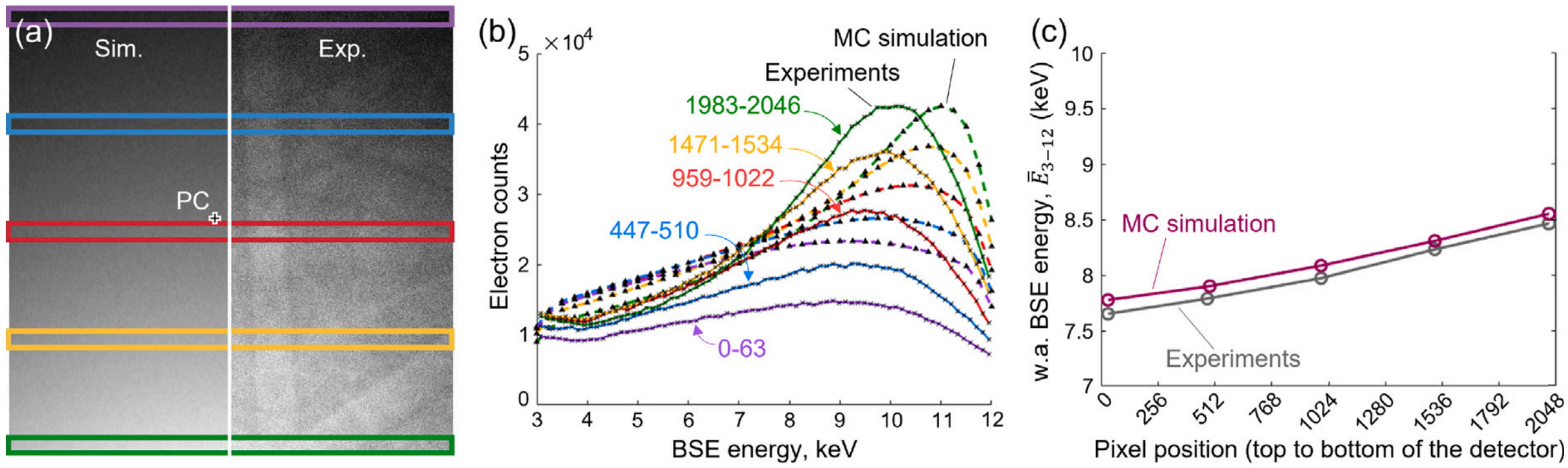

**Fig. 5.** (a) Left: Simulated cumulative BSE energy distribution on the detector using Monte Carlo simulations at 12 keV. Right: Experimental intensity map reported for comparison (replicated from Fig. 3c, but without color coding). (b) Simulated and experimental energy spectra within selected ROIs illustrated in (a). In this case, differently from Fig. 4c, the experimental electron count distribution is limited between 3 and 12 keV to match the simulated range of electron energy. (c) Comparison of simulated and experimental weighted average (w.a.) BSE energies, $\overline{E}$, (calculated from the 3–12 keV range), showing excellent agreement across all vertical positions.

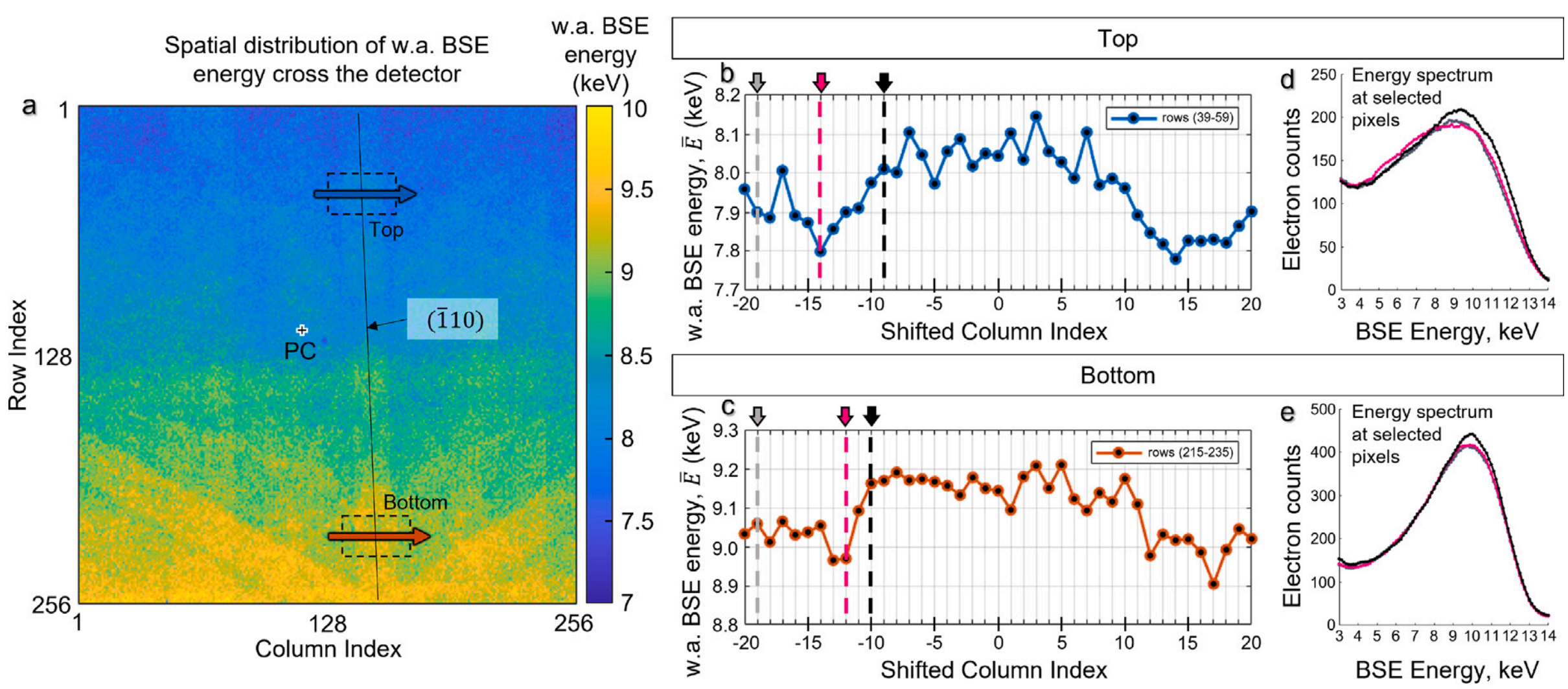

**Fig. 6.** (a) Weighted average (w.a.) BSE energy ($\overline{E}$) map on a pixel-by-pixel basis generated from the 12 keV dataset and extracted from the energy spectrum at each pixel, illustrating energy variations across the EBSD detector. A clear energy modulation is observed at the edges of Kikuchi bands, especially along the prominent vertical $(\bar{1}10)$ band. The signal in (a) should not be interpreted as a conventional EBSP or Kikuchi band. A slight gain variation is visible in the energy map in (a), which arises from the inability to acquire a flat gain reference within the SEM environment. (b, c) Averaged line profiles of $\overline{E}$ extracted from the top and bottom detector regions indicated in (a), revealing asymmetric energy decay from the band center to its edges. The line profiles are averaged over 20 pixels: (b) 39–59; (c) 215–235. The $x$-axis in (b) and (c) denotes the pixel column index across the detector, shifted such that the center of the Kikuchi band is aligned at pixel #0. (d, e) Energy spectra at selected pixel positions across the Kikuchi band indicated in (b,c) — gray: reference pixel; magenta: pixel displaying local minima of $\overline{E}$; black: pixel with high $\overline{E}$ along the line profile. These energy spectra demonstrate that shifts in low energy loss and high energy loss BSE counts, cause the changes in $\overline{E}$ observed in (b,c).

To elucidate the origin of this apparent energy decrease, we examine the BSE energy spectrum at selected pixel positions across the Kikuchi band (Figs. 6d and e). Column #-19, situated well outside the region of energy decay, serves as a baseline (gray spectrum in Figs. 6d and 6e). In the top region of the pattern in Fig. 6a, at map areas corresponding to local minima in $\overline{E}$, such as column #-14 (Fig. 6b), there is a relative increase in counts of low energy BSE (~3–8 keV, magenta spectrum in Fig. 6d). This localized enrichment in low energy BSEs directly accounts for the decrease observed in the $\overline{E}$ profile in Fig. 6b. Correspondingly, map areas with higher $\overline{E}$ (e.g., column #-9, Fig. 6b) exhibit an increase in counts of high energy BSE (~8–13 keV, black spectrum in Fig. 6d) versus the baseline. The spectrum from column #-14 (magenta spectrum) shows only slight deviation from baseline at high energy, and the spectrum from column #-9 (black spectrum) matches the baseline at low energy. Therefore, we conclude that in the top region of the detector, increases in $\overline{E}$ are caused by a localized increase in high energy electrons, while decreases in $\overline{E}$ are caused by a localized increase in low energy electrons.

For the bottom region of the pattern in Fig. 6a, the column with higher $\overline{E}$ (column #-10 in Fig. 6c, black spectrum in Fig. 6e) behaves similarly at high energies (~8–13 keV) to that of the top region (column #-9 in Fig. 6b, black spectrum in Fig. 6d). There is a marked increase in the intensity of high-energy electrons relative to the baseline (Fig. 6e), though the extra intensity in the bottom region is confined to a narrower energy range, ~9–11 keV. The column #-10 (black spectrum) matches closely with the baseline at low energies. Moreover, in contrast to the top region, the bottom region spectrum from areas of lower $\overline{E}$ (column #-12 in Fig. 6c, magenta spectrum in Fig. 6e) does not show an obvious increase in low energy electrons. There is also no decrease in high-energy electrons. Rather, the decrease in $\overline{E}$ comes from subtle changes in the overall spectrum counts versus the baseline.

Overall, the variations in the $\overline{E}$ are primarily governed by relative changes in the intensity of high-energy electrons (8–13 keV) near kikuchi bands, and vertically along the detector. At the top of the detector there is some contribution from relative changes in low-energy electrons.

It is also noteworthy that the absolute spectral intensity differs between the bottom and top regions of the detector (Figs. 6d and e). Although the number of low-energy electrons is comparable in both regions, the bottom of the detector records approximately twice as many high-energy electrons as the top. Consequently, even small variations in high-energy electron counts across pixels produce a more pronounced change in $\overline{E}$ in the bottom region, whereas the corresponding variation in the top region remains more gradual. This behavior accounts for the smooth decay of $\overline{E}$ observed in Fig. 6b and the sharper decline evident in Fig. 6c.

Taken together, the observations described so far also reflect the fact that high energy BSEs reach detector locations closer to the Kikuchi band centerline, while low energy electrons are detected farther from the center. This spatial modulation in electron counts originates from fundamental diffraction theory. For a given lattice spacing $d_{hkl}$, lower-energy electrons satisfy the Bragg condition at larger diffraction angles, leading to their detection at greater lateral distances from the band center. Conversely, higher-energy electrons, diffracted at smaller angles, arrive closer to the centerline. This diffraction-based behavior gives rise to the observed spatial variation in electron counts.

It is also important to understand the physical origin of the energy loss observed across Kikuchi bands, most notably along the {110} bands, where the variation of $\overline{E}$ at the band edges is particularly pronounced (see Fig. 6a). In particular, the $\overline{E}$ variation across Kikuchi bands underscores a non-trivial coupling between electron channeling and inelastic scattering processes arising from the electron–crystal interactions. A plausible explanation emerges from an analysis of the Kossel master pattern (simulated using EMsoftOO [45,47] by employing the Kossel pattern formation method described in [48]) as a function of depth within the sample, presented in Fig. 7, where successive patterns are computed in 5 nm increments, beginning at a depth of 5 nm. As depth increases, the intensity becomes progressively concentrated along the {110} bands and around < 110>-type zone axes, indicating enhanced electron channeling along these directions. In this context, it is proposed that electrons penetrating deeper into the sample are more likely to undergo inelastic processes prior to exiting the material [49]. This mechanism would explain the higher energy spread of electron energy across the {110} bands in Fig. 6a, as well as contribute to the asymmetric energy distribution observed across different Kikuchi bands.

Finally, while this may appear intuitive, it is critical to emphasize that all the interpretations of the data presented in this section reflect the fact that electrons of all energies in the distribution exit the sample carrying diffraction information and thus contribute to Kikuchi pattern formation. Indeed, if the spatially-dependent variations in electron counts across Kikuchi bands arise from fundamental diffraction behavior (as discussed above), it implies that even electrons with energies nearly half that of the primary beam participate in diffraction and carry crystallographic information, as further supported from the results reported in the next section.

### 3.3. Energy-filtered EBSD patterns

To quantitatively and visually evaluate the contribution of BSEs of varying energies to the overall, unfiltered BSE intensity, an EBSD dataset was acquired in integrating mode under the same experimental conditions described in Table 3 for the EBSD pattern shown in Fig. 4a, except that 100,000 sparse frames were collected instead of 1,000. By determining the energy of each detected electron, the individual frames were subsequently energy-filtered, enabling the reconstruction, through post-processing, of a series of EBSD patterns corresponding to discrete energy intervals and obtained under identical acquisition conditions. In this work, each energy-filtered EBSD pattern represents a 1 keV-wide interval within the 3–12 keV range (i.e., 3–4 keV, 4–5 keV, . . . , 11–12 keV). The high number of collected frames ensured that a sufficient number of BSEs were detected in each energy interval, thereby providing adequate SNRs for the reconstruction and comparison of EBSD patterns across all energy windows, without limitations arising from noise or variations in exposure time. A direct comparison of the resulting EBSD patterns is shown in Fig. 8. Fig. 8a illustrates the distribution of electron counts derived from the unfiltered EBSD pattern, while representative examples of energy-filtered EBSD patterns (reconstructed by applying discrete energy windows to the unfiltered dataset) are displayed in Fig. 8b. All 1 keV-wide energy-filtered EBSD patterns obtained within the 3–12 keV range are reported in Figure S.3a of the Supplementary Material. To facilitate a meaningful visual comparison, the intensities of all energy-filtered patterns have been normalized to a common scale (excluding the unfiltered reference). Fig. 8c presents the total intensity of each energy-filtered EBSD pattern (computed as the integral of the distribution in Fig. 8a over the corresponding energy interval), normalized by the total intensity of the 3–4 keV EBSD pattern. This provides a direct estimate of the relative number of BSEs in each energy interval, thereby quantifying the contribution of BSEs with different energies to the final, unfiltered BSE intensity, as would be observed in a conventional EBSD measurement. Fig. 8c shows that electrons with an energy loss of approximately 15%–25% of the primary beam energy constitute the main contributors to EBSD pattern formation, exhibiting a normalized total intensity about 3.5 times higher than that of 3–4 keV BSEs. Notably, electrons in the 3–8 keV range also provide a significant contribution to the overall diffraction signal.

To isolate the diffraction contribution and suppress background effects, each energy-filtered pattern was gain-corrected using a reference background, also energy-filtered for each energy interval, acquired from an amorphous Si region located on the same specimen. The gain reference was obtained by collecting 100,000 sparse frames under identical working distance, camera length, and acquisition conditions as those used for the Si (100) EBSD measurements, thereby ensuring experimental consistency. Fig. 8d shows a direct comparison of selected gain-corrected, energy-filtered EBSD patterns, with all intensities normalized to a common scale for meaningful visual comparison (excluding the unfiltered reference). The complete set of 1 keV-wide gain-corrected energy-filtered EBSD patterns within the 3–12 keV range is provided in Figure S.3b of the Supplementary Material. This analysis reveals that high-loss electrons (energy loss > 4 keV) carry diffraction information and therefore contribute to the EBSD Kikuchi signal.

As a complementary approach to this analysis, a second experiment was conducted to examine EBSD patterns acquired across different energy-filtering intervals under comparable signal conditions.

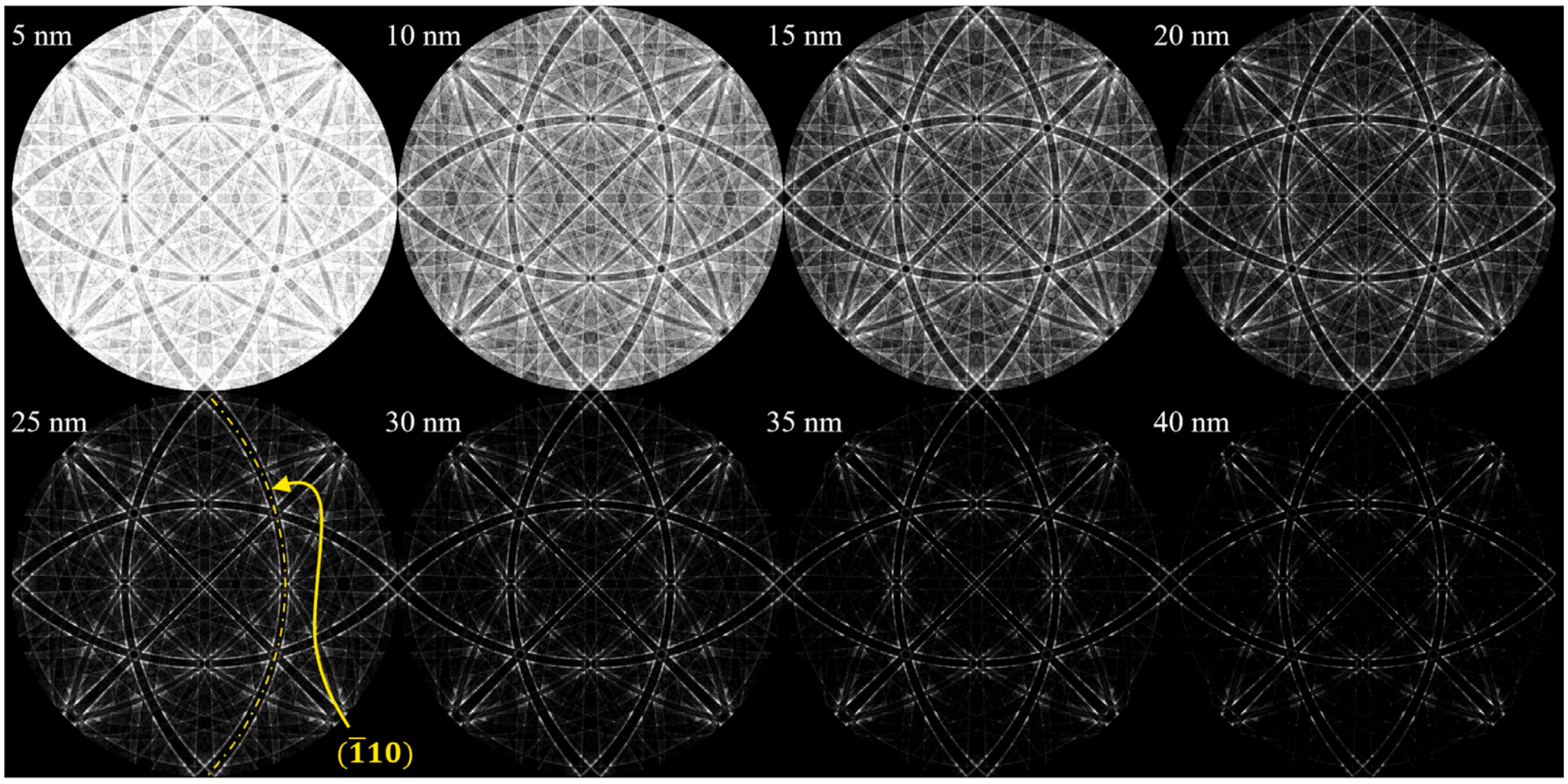

**Fig. 7.** Depth-resolved simulation of the Kossel master pattern, computed in 5 nm increments beginning at a depth of 5 nm. The results reveal a progressive concentration of intensity along the {110} Kikuchi bands and around < 110>-type zone axes with increasing electron penetration depth.

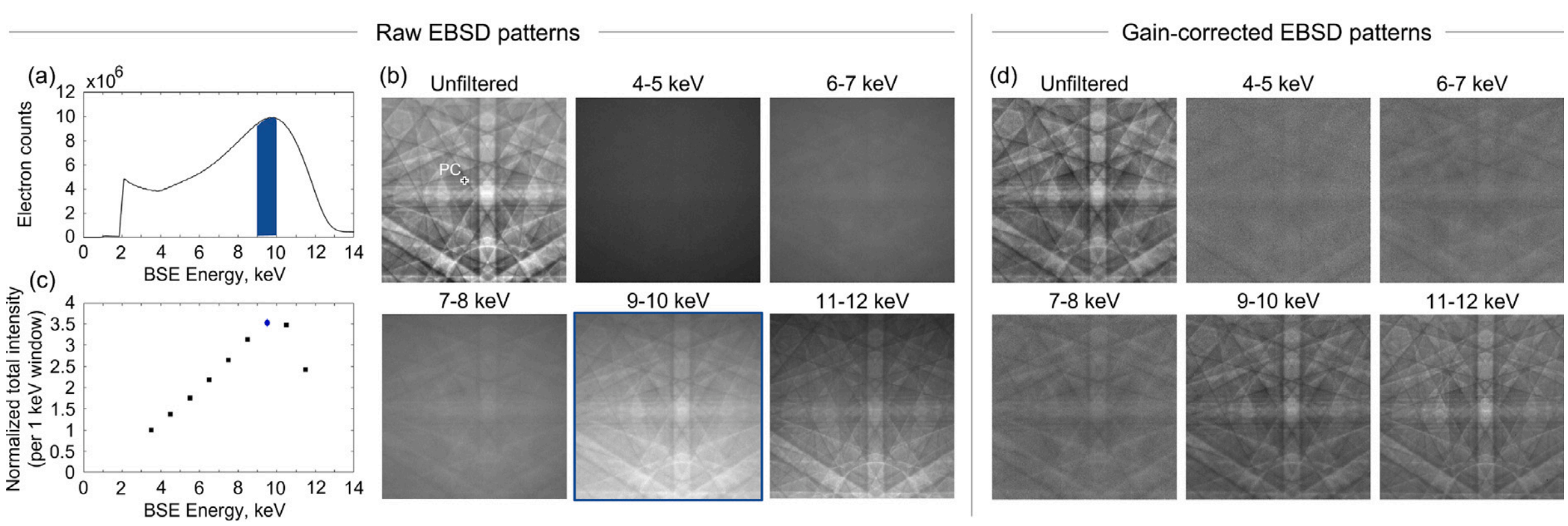

**Fig. 8.** (a) Distribution of electron counts derived from the unfiltered EBSD pattern in (b). (b) Unfiltered EBSD pattern as well as representative energy-filtered EBSD patterns reconstructed from 1 keV-wide energy intervals between 3 keV and 12 keV. The energy-filtered EBSD patterns are normalized to a common intensity scale. (c) Total integrated intensity of each energy-filtered EBSD pattern, normalized to the 3–4 keV interval, providing the relative number of BSEs in each energy range. (d) Corresponding gain-corrected energy-filtered EBSD patterns, normalized to a common intensity scale (except for the unfiltered pattern) for direct visual comparison.

This analysis was intended to address the difficulty of directly comparing energy-filtered patterns with inherently different signal intensities, thereby enabling a further quantitative evaluation of the energy-resolved EBSD data. For this, single EBSD patterns were reconstructed using a 12 keV primary beam by summing sparse frames collected with energy-selective filtering of the detected electrons over three distinct energy intervals: 2–8 keV, 8–10 keV, and 10–13 keV. Importantly, all patterns were acquired in counting mode, which ensures equal weighting of electrons of different energies. Because the number of detected electrons per pixel per frame varies with the selected energy range, the number of frames summed in each case, and consequently the exposure time, was adjusted to yield an average signal intensity of approximately 2,000 electrons per pixel in the final pattern. This ensured a good SNR so that diffraction signal quality could be evaluated [4]. The corresponding exposure times are reported in Fig. 9 and reached up to 16 minutes per pattern due to the extremely low beam currents required for resolving individual BSE events on the detector. The variation in exposure time between energy intervals reflects the changes in detected BSE intensity with energy, consistent with the distribution shown in Fig. 8a. The resulting energy-filtered EBSD patterns, along with an unfiltered reference acquired in both integrating and counting modes, are presented in Fig. 9. Note that for the unfiltered integrating-mode EBSD pattern, the number of summed frames is the same as that used in unfiltered counting-mode to ensure fair comparison between acquisition modalities.

Fig. 9 shows that, again, despite the use of a nominal 12 keV primary beam, electrons in the markedly lower 2–8 keV range still generate diffraction patterns with clear and well-defined features. This observation is particularly significant, as it provides direct evidence that diffraction contrast is not governed exclusively by electrons near the incident beam energy, but also by those that have undergone substantial energy loss (further considerations are presented in Appendix B). This finding offers new insight into the long-standing question of the energy thresholds required for EBSD signal generation. Another interesting feature of the 2–8 keV filtered EBSP is the contrast inversion of the higher-order Laue zone (HOLZ) in the lower half of the pattern, a feature that is further elucidated in the following section.

[4] DED images are generally limited by Poisson noise from the electron gun. The SNR is $\sqrt{N}$, where $N$ denotes the number of electrons (signal) [50]. Typically, a good SNR is above 30, corresponding to approximately 1000 electrons per pixel. The SNR for Fig. 9 is ~45.

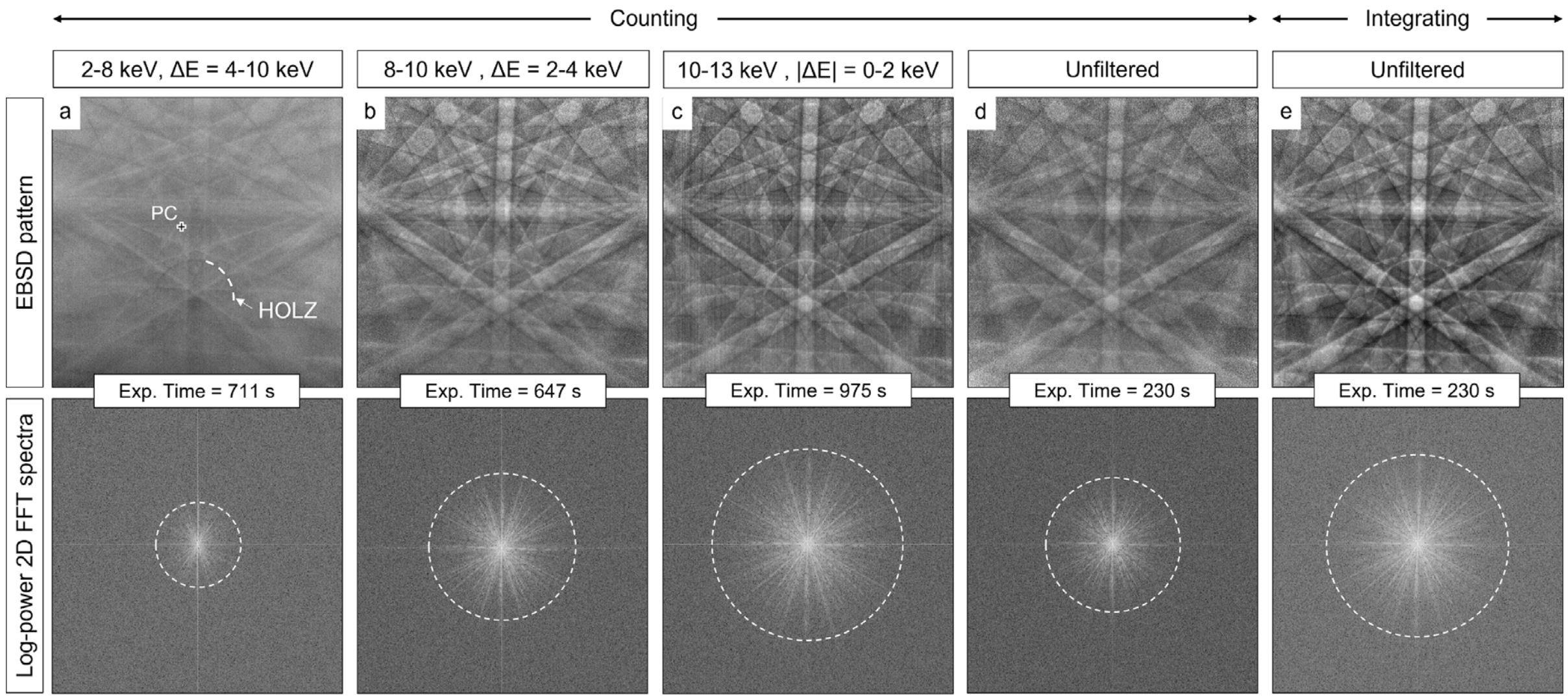

**Fig. 9.** Top: Raw EBSD patterns acquired at 12 keV primary beam energy using (a–c) energy-filtered counting mode for three energy intervals (2–8, 8–10, and 10–13 keV), and unfiltered patterns obtained in (d) counting and (e) integrating modes. Note that (d) should not be interpreted as the average of (a), (b), and (c), as electrons with energies above 13 keV, arising mainly from multi-pixel events, are detected in the unfiltered counting mode. Only the same dark reference background is applied to all the EBSD patterns, which is acquired with the detector positioned in the chamber, the electron beam blanked, and all photon sources inactive. All patterns were normalized to an average of 2000 electrons per pixel by adjusting the number of summed frames to ensure comparable SNRs. The unfiltered integrating-mode pattern uses the same number of frames as the unfiltered counting-mode case. Bottom: corresponding log-power 2D FFT spectra.

To further analyze the impact of electron energy on diffraction contrast, the spatial frequency content of the energy-filtered patterns is examined through their log-power 2D fast Fourier transform (FFT) spectra. In such transforms, low spatial frequencies (associated with broad intensity variations) appear near the center of the spectrum, while high spatial frequencies (reflecting fine detail and sharp transitions) occupy the outer regions. The filtered pattern generated by 2–8 keV electrons exhibits minimal high-frequency content, consistent with its smoother and more diffuse appearance (Fig. 9a). In contrast, the 10–13 keV image displays a significantly richer high-frequency spectrum, indicative of sharp Kikuchi band edges and finer crystallographic detail (Fig. 9c). These observations clearly support the view that low-energy-loss electrons contribute more prominently to Kikuchi pattern formation, consistent with numerous prior studies [5,7–9], and that filtering high-energy-loss electrons improves Kikuchi band contrast by reducing diffuse background contributions [16,23,24]. However, at the same time, this does not exclude a meaningful contribution to EBSD pattern formation from electrons that have undergone higher energy losses.

When compared to the unfiltered counting-mode image (Fig. 9d), the increase in frequency content of 10–13 keV filtered patterns translates into noticeably enhanced sharpness, underscoring the role of energy selection in improving pattern resolution. Notably, the unfiltered integrating-mode pattern also exhibits a comparably rich high-frequency spectrum (Fig. 9e), closely resembling that of the 10–13 keV energy-filtered counting-mode EBSD pattern (Fig. 9c). At first glance, it may appear counterintuitive that unfiltered patterns acquired in counting mode suffer in high-frequency content. However, this observation aligns with the underlying image formation mechanism. In counting mode, each detected electron contributes equally to the image, regardless of its energy, allowing low-energy electrons to blur the resulting contrast. Conversely, integrating mode inherently weights electrons by their detected signal, which is approximately proportional to their energy. This means low-energy electrons contribute less strongly to image contrast. Energy-filtered counting mode further refines this approach: by restricting detection to electrons within a narrow keV window centered on the primary beam energy, it enhances the dominance of high-energy electrons in contrast formation. As a result, the log-power spectra of the corresponding patterns show the highest high-frequency content for the 10–13 keV energy-filtered case, followed by integrating mode, and finally counting mode.

It is also important to consider that the relative performance of counting versus integrating mode depends significantly on the MTF, or equivalently, the DQE at higher spatial frequencies. For the present detector, the average spatial spread of the electron signal is 1.4 pixels (see Figure S.2 in the Supplementary Materials), indicating a high MTF in integrating mode. However, for detectors or settings with reduced MTF, integrating mode may no longer offer an advantage over counting mode.

## 4. Discussion and outlook

In discussing the energy spectra measured from EBSPs, it is useful to think not only in terms of high and low energy, but also in terms of energy-loss. High energy electrons can be taken as a synonym for low-energy-loss or quasi-elastically scattered electrons. These are known as low-loss electrons in electron energy loss spectroscopy. We can define the change in BSE energy $\Delta E$ as $\Delta E = E_{\text{beam}} - E_{\text{BSE}}$, where $E_{\text{beam}}$ is the primary beam energy and $E_{\text{BSE}}$ is the measured energy of an individual electron.

Numerous studies have underscored the dominant role of high energy electrons (i.e., those retaining a significant fraction of the primary beam energy) in generating high-contrast diffraction features [5,7–9]. Energy-filtered EBSD experiments, in particular, have shown that pattern clarity often peaks when electrons within approximately 1%–3% of the incident energy are selectively imaged, with effective energy windows frequently centered within 0.5–1.5 keV of the primary beam energy [7]. Other studies have come to the conclusion that the “effective Kikuchi pattern spectrum” is characterized by a narrow energy distribution, typically with $\Delta E$ less than 1 keV, peaking slightly below the primary beam energy [5,8,9,13]. At the same time, however, Deal et al. [7] employed an energy filtering window (not threshold) centered at 80% of the 15 keV primary beam energy (i.e., 11.5-12 keV) during the acquisition of EBSD patterns from Si, Fe, and Ir, demonstrating the clear presence of Kikuchi bands within this range. In this context, our results demonstrate that the dominant contribution to EBSD pattern formation, corresponding to the peak in electron counts, originates from electrons that have lost 15%–25% of the primary beam energy (Fig. 4b, Fig. 5b, and Fig. 8). Furthermore, electrons exiting the sample with energies in the 3–8 keV range are found to retain substantial diffraction information and contribute significantly to the formation of EBSD patterns, as illustrated in Fig. 8 and Fig. 9a.

The variation of BSE energy as a function of scattering angle has also received considerable attention, particularly in light of its implications for pattern geometry and resolution. For example, Ram and De Graef [12], using MC simulations that incorporate the CSDA to describe inelastic scattering, reported a several-keV energy gradient across the detector, in line with what observed in this work (Fig. 5c), where the weighted average BSE energy increases toward the bottom of the detector. It was therefore recommended that this energy variation be incorporated into all EBSD pattern simulation methodologies. In a subsequent study, Winkelmann et al. [13] conducted an energy-resolved pattern matching analysis, comparing experimental EBSD data with simulated patterns. Their findings indicated a comparatively narrower energy distribution of BSEs (approximately 1 to 1.5 keV of $\Delta E$) contributing to diffraction contrast, along with a more moderate dependence of electron energy on scattering angle.

Further insight into the nature of electron energies contributing to Kikuchi pattern formation has been advanced through Winkelmann's modeling framework [14]. In the conventional understanding [1], Kikuchi patterns arise through a two-step process: initially, electrons undergo inelastic, incoherent scattering events (such as phonon excitation and electronic processes including plasmon and core-level losses) that reduce their energy. Subsequently, a subset of these electrons experiences coherent diffraction within the crystal, giving rise to Kikuchi patterns. Winkelmann's approach builds on this foundation by emphasizing that high-contrast diffraction features predominantly arise from high energy (low $\Delta E$) electrons, as those experiencing significant inelastic scattering (higher $\Delta E$), particularly delocalized plasmon scattering, tend to lose diffraction contrast. However, the model does not exclude that electrons with moderate energy losses may still contribute meaningfully. Through anomalous absorption mechanisms dependent on crystal structure, some electrons scattered incoherently are redirected and re-localized, effectively acting as new point sources for Kikuchi pattern formation [14]. This phenomenon broadens the energy range of electrons contributing to the observed diffraction signal beyond the narrow high energy (low $\Delta E$) regime. Fig. 9 provides experimental evidence to support this theory.

In this context, by combining direct electron detection with pixel-wise energy mapping, the present study offers new experimental access to the spatially resolved energy distribution of BSE contributing to EBSD patterns. Our experimental results reveal both a measurable angular gradient in weighted average BSE energy across the detector and that the spectrum of electrons contributing to EBSD pattern formation extends more broadly, encompassing not only high-energy (low $\Delta E$) electrons but also those with substantial inelastic losses (higher $\Delta E$) that still carry meaningful crystallographic information, albeit with lower contrast.

The contrast inversion observed in the region below the pattern center in Fig. 9a is also particularly striking. Several factors may contribute to this behavior. For instance, since HOLZ features originate predominantly from near-elastic backscattered electrons, applying an energy filter that selects only electrons in the 2–8 keV range (corresponding to large energy losses) substantially suppresses their contribution, thereby reducing the visibility of these fine diffraction structures. In addition, the contrast inversion likely arises from the complex interplay between inelastic scattering and depth-dependent diffraction contrast. Specifically, Winkelmann [15] reported a similar inversion behavior in depth-resolved EBSD simulations, where patterns generated from deeper crystal slices displayed reversed contrast relative to those near the surface, despite their comparatively smaller contribution to the overall intensity. By analogy, it can be posited that energy filtering represents a form of depth-resolved EBSD, in which higher-$\Delta E$ electrons, having undergone more extensive scattering, predominantly originate from greater depths within the interaction volume. The observed contrast inversion in the high-$\Delta E$ regime may thus reflect a depth-dependent reversal in diffraction conditions, rather than being solely attributed to elastic or compositional effects.

Moreover, the implementation of single-electron-level energy-selective filtering windows introduced in this study offers a powerful advancement for EBSD, enabling the targeted isolation of electron populations predominantly undergoing elastic scattering. Unlike previous studies in which a qualitative assessment of electron energy measurements is used to define energy thresholds and therefore allow filtering during acquisition [7,16,17,21–25], this work presents the first methodology for quantitative measurement of electron energy within EBSD, enabling the precise selection of energy windows tailored to enhance pattern quality. Such flexibility in selectivity has the potential to produce sharper Kikuchi bands, reduced background noise, and improved overall pattern fidelity. Additionally, the possibility of applying the electron counting algorithm *ex post facto* to sparse EBSD frames originally acquired in integrating mode further extends the scope of energy-resolved post-processing. These enhancements are particularly critical in high-resolution EBSD (HR–EBSD), where the accuracy of strain and rotation measurements relies on the precise definition of band edges and the detection of subtle shifts in their positions. By sharpening these features through energy filtering, the technique not only enhances angular resolution but also significantly improves the reliability and sensitivity of EBSD in resolving lattice distortions and distinguishing crystallographically similar phases. Lateral resolution would also benefit from the combined use of energy filtering and lower accelerating voltages, which reduce the interaction volume and limit contributions from deeper scattering.

Finally, it is critical to acknowledge that, although high energy electrons (low $\Delta E$) offer the highest fidelity in crystallographic information, the primary constraint in most EBSD measurements remains the SNR. Enhancing SNR is intrinsically linked to the number of detected electrons; thus, any form of energy-based electron filtering must be approached with caution. Electrons that have undergone significant energy loss, while potentially reducing fine-detail clarity in the EBSP, still contribute meaningfully to the visibility of low-frequency features such as Kikuchi bands. Accordingly, energy filtering can be detrimental to the overall pattern quality unless the unfiltered pattern already exhibits a high SNR. As such, we do not anticipate a benefit of energy filtering for standard orientation mapping. However, in the context of HR–EBSD, where prolonged integration times yield the requisite SNR, selective energy filtering is expected to offer meaningful improvements.

## 5. Conclusions

This study presents a major advancement in EBSD by introducing a fully energy-resolved detection approach capable of measuring the energy of every electron contributing to diffraction pattern formation. Enabled by the use of a MAPS-based DED, a centroid-based electron counting algorithm, and supported by a robust calibration method, this approach reconstructs the energy spectra of backscattered electrons with spatial resolution across the detector. The results offer a step forward from conventional EBSD, which integrates over energy, to a method that captures both the intensity and spectral composition of electron scattering events. The main findings of this work are as follows:

- An energy-resolved EBSD acquisition framework was implemented using MAPS-based detectors in integrating mode, with electron energies extracted *ex post facto* via centroid-based clustering of sparse signals. Calibration in direct beam mode across multiple beam energies enabled accurate conversion of ADU to energy, revealing both a ~ 8 ADU/keV conversion factor and a ~ 1 keV energy resolution for our DE-SEMCam detector.

- Using the calibrated ADU/keV conversion, the centroid-based algorithm was applied to sparsely populated EBSD frames on Si(100), yielding a detailed BSE energy spectrum. Despite a 12 keV primary beam, inelastic scattering broadened the energy down to 3 keV ($\Delta E = 9$ keV), with a vertical gradient of weighted average electron energy, decreasing from bottom to top of the detector, closely matching Monte Carlo predictions of spatially varying escape behavior.
- The analysis of the spatial distribution of BSE energy on the detector revealed systematic variations in the weighted average electron energy at Kikuchi band edges. These variations reflect a broadened energy distribution, particularly along {110} bands and <110> zone axes, where deeper electron channeling likely enhances inelastic scattering and contributes to the observed broadening.
- Energy-resolved reconstruction of 1 keV-wide EBSD patterns revealed that electrons exhibiting an energy loss of approximately 15%–25% of the primary beam energy dominate Kikuchi pattern formation. The distribution of electron counts across the energy intervals directly quantifies the relative contribution of BSEs to the overall, unfiltered EBSD intensity. Gain-corrected analysis confirmed that even electrons with larger energy losses ($\Delta E \geq 4$ keV) retain diffraction contrast. These results demonstrate that a broad range of backscattered electrons, including those significantly inelastically scattered, contribute meaningfully to EBSD pattern generation.
- A comparison between acquisition modes (integrating vs. counting) revealed fundamental differences in contrast formation: the integrating mode favors high-energy (low $\Delta E$) electrons due to signal-weighting, while counting mode gives equal weight to all. When combined with energy filtering around the primary beam energy ($|\Delta E| = 0$–$2$ keV), counting mode yielded the sharpest patterns and the richest high-frequency detail, reflecting the enhanced contribution of elastically scattered (low $\Delta E$) electrons.

In conclusion, this work elevates EBSD from a pattern-matching technique to a quantitative spectro-crystallographic tool. The ability to directly measure and filter the energy contributions of BSE to pattern formation not only deepens the understanding of diffraction contrast mechanisms but also introduces a versatile tool-set for improving pattern quality and expanding quantitative capabilities. As energy-resolved EBSD continues to evolve, it promises to enhance theoretical modeling, guide experimental design, and enable more precise crystallographic characterization across a broad spectrum of materials science applications.

## CRediT authorship contribution statement

**Nicolò M. Della Ventura:** Writing – original draft, Visualization, Software, Methodology, Investigation, Formal analysis, Data curation, Conceptualization. **Kalani Moore:** Writing – original draft, Visualization, Software, Methodology, Investigation, Formal analysis, Data curation, Conceptualization. **McLean P. Echlin:** Writing – original draft, Visualization, Software, Methodology, Investigation, Formal analysis, Data curation, Conceptualization. **Matthew R. Begley:** Writing – review & editing, Supervision, Resources, Project administration, Funding acquisition. **Tresa M. Pollock:** Writing – review & editing, Supervision, Resources, Project administration, Funding acquisition. **Marc De Graef:** Writing – review & editing, Visualization, Software, Methodology, Investigation, Formal analysis, Data curation, Conceptualization. **Daniel S. Gianola:** Writing – review & editing, Supervision, Resources, Project administration, Methodology, Funding acquisition, Formal analysis, Conceptualization.

## Declaration of competing interest

The authors declare that they have no known competing financial interests or personal relationships that could have appeared to influence the work reported in this paper.

## Acknowledgments

This research was supported by funds from the UC National Laboratory Fees Research Program of the University of California, United States, Grant Number L22CR4520, and from the Army Research Laboratory, United States accomplished under Cooperative Agreement Number W911NF-22-2-0121. The views and conclusions contained in this document are those of the authors and should not be interpreted as representing the official policies, either expressed or implied, of the Army Research Laboratory or the U.S. Government. The U.S. Government is authorized to reproduce and distribute reprints for Government purposes notwithstanding any copyright notation herein. Direct Electron's work in this publication was supported by the National Institute of Mental Health (NIMH) of the National Institutes of Health (NIH), United States under award number 1R44MH125687. The authors also acknowledge the NSF MRI instrumentation grant No. 2117843. The research reported here made use of the shared facilities of the Materials Research Science and Engineering Center (MRSEC) at UC Santa Barbara: NSF DMR–2308708. The UC Santa Barbara MRSEC is a member of the Materials Research Facilities Network (www.mrfn.org). MDG acknowledges financial support from a National Science Foundation, United States grant (DMR-2203378), use of the computational resources of the Materials Characterization Facility at Carnegie Mellon University, United States supported by grant MCF-677785, as well as support from the John and Claire Bertucci Distinguished Professorship in Engineering. The authors gratefully acknowledge F. Wang and A. Taylor's early contributions, which, although not reflected in the data presented here, were instrumental in initiating this study.

## Appendix A. Energy deposition in DE-SEMCam epitaxial layer

See Fig. A.1.

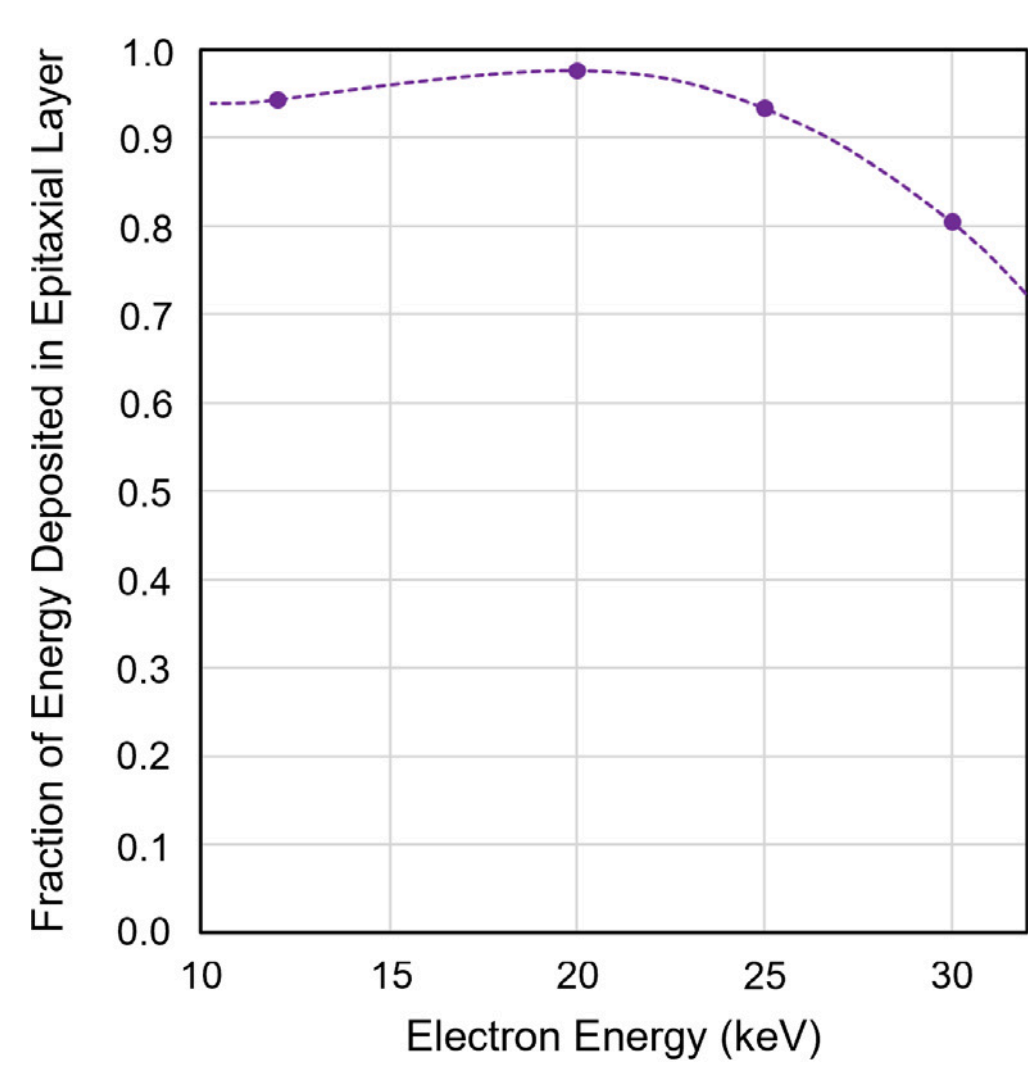

**Fig. A.1.** The simulated fraction of incident electron energy deposited in the epitaxial layer of the DE-SEMCam detector at various energies. Simulations were completed using PENELOPE [51].

## Appendix B. Further considerations on the 2–8 keV energy-filtered EBSD pattern

Here, we present complementary arguments confirming that, under a 12 keV primary beam, electrons with substantial energy loss (as those in the 2–8 keV range, $\Delta E$ = 4–10 keV) also contribute to diffraction pattern formation. Specifically, when restricting the energy loss $\Delta E$ of BSEs that contribute to the formation of the Kikuchi pattern to only a maximum of 2 keV from the 12 keV primary beam (that is, considering only electrons with at least 10 keV residual energy), the detector's resulting ADU histogram (which accounts for the detector's 1 keV energy resolution) would align closely with that obtained during the calibration of the detector at 10 keV, exhibiting ADU values predominantly in the 65–95 range (Fig. 2). Applying the established ADU/keV conversion factor for 10 keV electrons (8.11 ADU/keV - Table 2) yields a minimum detected energy of approximately 8.05 keV. Notably, this value is above the 2–8 keV energy window used in the filtered diffraction dataset, and electrons at this energy level would therefore be excluded from detection, contributing no signal to the filtered pattern. Consequently, even electrons that have lost only up to 2 keV cannot account for the diffraction features observed in the 2–8 keV filtered patterns, also when the finite energy resolution of the detector is taken into consideration. This reinforces the interpretation that electrons undergoing large energy losses do play a role in generating the Kikuchi pattern.

## Appendix C. Supplementary data

Supplementary material related to this article can be found online at https://doi.org/10.1016/j.ultramic.2025.114301.

## Data availability

Data will be made available on request.